\documentclass{eas}
\usepackage{graphicx}
\begin{document}

\title{Status of R\&D on Micromegas for Rare Event Searches: The T-REX project}
\runningtitle{Irastorza \etal: T-REX project}

\author{I~G~Irastorza$^{1}$, J~Castel$^{1}$, S~Cebri\'an$^{1}$, T~Dafni$^{1}$, G~Fanourakis$^{3}$, E~Ferrer-Ribas$^{2}$, D~Fortu\~{n}o$^{1}$, L~Esteban$^{1}$,
J~Gal\'an$^{2}$, J~A~Garc\'ia$^{1}$, A~Gardikiotis$^{4}$, J~G~Garza$^{1}$, T~Geralis$^{3}$, I~Giomataris$^{2}$, H~G\'omez$^{1}$, D~C~Herrera$^{1}$,
F~J~Iguaz$^{2}$, G~Luz\'on$^{1}$, J~P~Mols$^{2}$, A~Ortiz$^{1}$, T~Papaevangelou$^{2}$,
A~Rodr\'iguez$^{1}$, J~Ruz$^{5}$, L~Segui$^{1}$, A~Tom\'as$^{1}$, T~Vafeiadis$^{6,5,4}$ and S~C~Yildiz$^{7,}$}

\address{Grupo de F\'isica Nuclear y Astropart\'iculas, University of Zaragoza, Zaragoza, Spain}
\address{IRFU, Centre d' \'Etudes de Saclay, CEA, Gif-sur-Yvette, France}
\address{Institute of Nuclear Physics, NCSR Demokritos, Athens, Greece}
\address{University of Patras, Patras, Greece}
\address{CERN, European Organization for Particle Physics and Nuclear Research, Switzerland}
\address{Aristotle University of Thessaloniki, Thessaloniki, Greece}
\address{Bo\u{g}azi\c{c}i University, Istanbul, Turkey}
\address{Do\u{g}u\c{s} University, Istanbul, Turkey}

\begin{abstract}
The T-REX project aims at developing novel readout techniques for Time Projection Chambers in experiments searching for rare events.
The enhanced performance of the latest Micromegas readouts in issues like energy resolution,
gain stability, homogeneity, material budget, combined with low background techniques, is opening new windows of opportunity for their application in this field. Here we review the latest results regarding the use and prospects of Micromegas readouts in axion physics (CAST and the future helioscope), as well as the R\&D carried out within NEXT, to search for the neutrinoless double-beta decay.

\end{abstract}
\maketitle
\section{The T-REX project}

The common characteristic of the so-called rare event searches, like axion, dark matter or double-beta decay
($\beta\beta$) searches, is the extremely low signal rate expected. The ultra-low backgrounds required are achieved by the use of active and passive shielding, operation in underground sites, event discrimination and a careful selection of the detector
materials from the radiopurity point of view.

Gaseous time projection chambers (TPC) can offer features of particular interest for rare event detection. The topological information of the event in gas, precisely registered by an appropriately patterned readout, is a powerful tool for signal identification and background rejection. However, only recent advances in TPC readouts based on micropattern gas detectors (MPGD) have allowed the competitive proposal of gas TPCs in different rare event applications. The technological boost provided by these readouts
relies in the use of metallic strips or pads, precisely printed on plastic supports with photolithography
techniques, substituting the conventional multiwire proportional chambers (MWPC). The simplicity,
robustness and mechanical precision are much higher than those of MWPCs. One of the most
attractive MPGD for application in rare events, is the Micromegas readout plane (Giomataris {\em et al.\/} \cite{mm}).

The T-REX project aims at exploring the latest gas TPC readout concepts and merging them with low background know-how. The main component of these studies focuses on the latest generation Micromegas \textit{microbulk} readout planes and their further development in order to meet different requirements. We present here two specific applications in which much activity is going on lately, the Micromegas detectors for low background x-ray detection in the CAST experiment (CAST collaboration \cite{cast1,cast2,cast3}) for axion searches, and the studies of Micromegas readouts for double beta decay within the NEXT project (NEXT collaboration \cite{loi,cdr}).

\section{Micromegas in the search for solar axions}

The CAST (CERN Axion Solar Telescope) experiment is the most powerful implementation of the ``helioscope concept'' searching for hypothetical axions coming from the Sun. The experiment is counting with a powerful dipole magnet that triggers the conversion of the axions into photons of energy 1-10 keV. Low background x-ray detectors are therefore necessary for high sensitivity. Recently, a much improved version of the helioscope concept has been proposed (Irastorza {\em et al.\/} \cite{ngah}) exploiting the innovations introduced by CAST and improve its sensitivity substantially. One of the necessary elements of this new generation axion helioscope are x-ray detectors with even lower backgrounds. This is the main motivation of the development here presented.

Micromegas detectors are being used in CAST since the beginning of the experiment in 2003.
CAST has been a test ground for these detectors, where they have been combined with low background techniques like the
use of shielding, radiopurity screening of detector components or advanced event discrimination techniques based on the detailed topological information offered by the readout. The CAST detectors were the first Micromegas readouts with a 2D pattern, with a pitch of 300 microns (Abbon {\em et al.\/} \cite{Abbon:2007ug}).

\begin{figure}[htb!]
\begin{center}
\includegraphics[width=16pc]{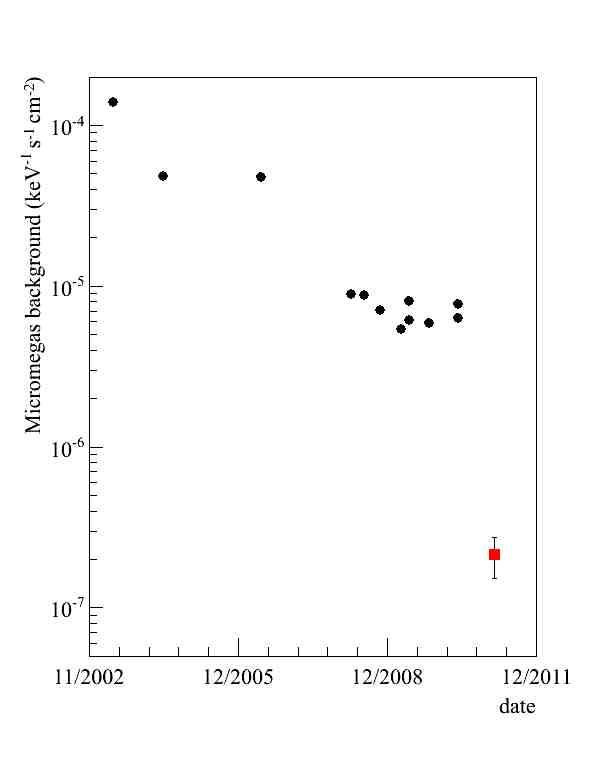}\hspace{2pc}%
\begin{minipage}[b]{11pc}\caption{\label{history} A history of the background levels of the Micromegas detectors
registered in different periods of data taking in the CAST experiment. The
last point on the right represents the background level recorded at the LSC under special shielding conditions (see text).}
\end{minipage}
\end{center}
\end{figure}

Since the upgrade in 2007, the number of Micromegas installed in the experiment was increased from one to three, being now all three of them of the microbulk type (Andriamonje {\em et al.\/} \cite{mbulk}). During the upgrade there were improvements in the shielding of all detectors, which in combination with the better characteristics of the microbulks, has lead to an improvement of the detector performance. The effect on the background
levels can be appreciated in the plot of Figure \ref{history}: they have dropped by at least a factor of 20, with the
current detectors showing levels of 6--9$\times$10$^{-6}$s$^{-1}$cm$^{-2}$keV$^{-1}$.

Current efforts are focused on understanding the sources of the present background level and design strategies to further reduce it in subsequent designs of the detectors setup. A copy of one of the CAST detector systems was prepared and installed to take data in the Zaragoza laboratory. The setup
is housed inside a Faraday cage, with an inner shielding of copper and lead, an outer shielding of
polyethylene and it has been completed with a nitrogen flux in the vicinity of the chamber, reproducing
the setup of the CAST detectors (Gal\'an {\em et al.\/} \cite{Gal10}; Dafni {\em et al.\/} \cite{Daf11}).

\begin{figure}[htb!]
\begin{center}
\includegraphics[width=15pc]{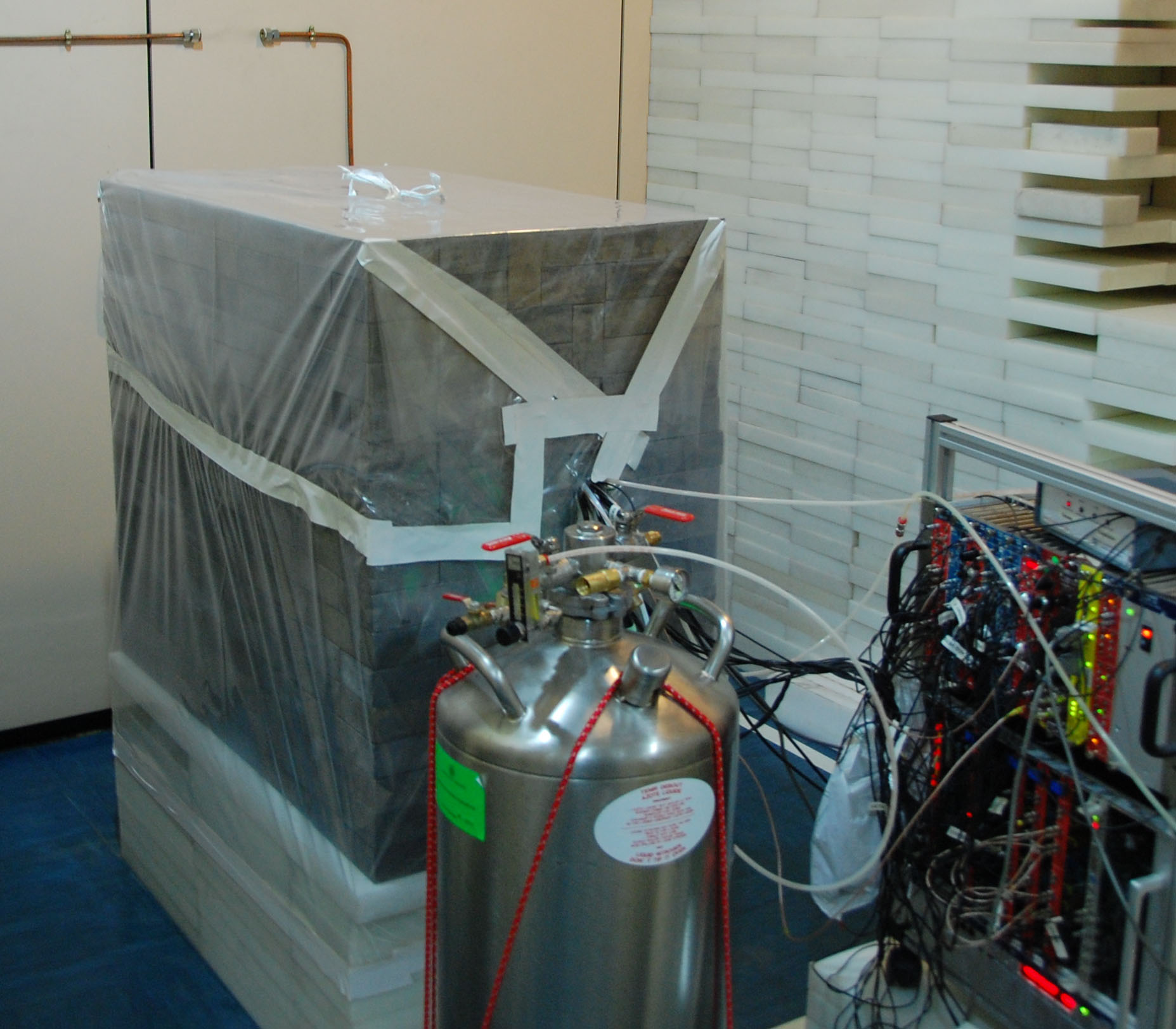}\hspace{2pc}%
\begin{minipage}[b]{12pc}\caption{\label{lsc1} A photo of the setup in the LSC:
The Faraday cage that houses the detector is placed inside a 4$\pi$ lead shielding.
The dewar placed in front of the shielding provides the Nitrogen flux.}
\end{minipage}
\end{center}
\end{figure}

After a short period of data-taking, the system was moved underground, in the Canfranc Underground Laboratory
(LSC) in the Spanish Pyrenees. In this environment data have been taken in different shielding configurations, and in well known conditions regarding external gamma background to assure a reliable comparison with simulation data. A continuous Nitrogen flux is provided to purge the radon that could stay close to the detector (Figure \ref{lsc1}). A detailed account of the tests performed is presented elsewhere (Tom\'as {\em et al.\/} \cite{alfredo}, Gal\'an {\em et al.\/} \cite{Gal11}).

Although the studies are still ongoing, levels as low as  ~2$\times$10$^{-7}$s$^{-1}$cm$^{-2}$keV$^{-1}$ have already been reached in the configuration with the largest lead thickness (20 cm), a level much lower than the one at CAST (Figure \ref{history}). The preliminary conclusions are that this is the intrinsic level of the radiopurity of the detector components, while the level obtained at CAST is still dominated by external gammas, probably from the poorly shielded solid angle due to the detector's connection to the magnet.
These studies have provided the needed information in order to
design improvements on the CAST shieldings that are being implemented for the next data taking phase. As a subsequent step a new design of the detector is being carried out to better incorporate these learnings. The work at LSC is now focused on studying and identifying internal sources of background, whose removal could eventually lead to improvements beyond the 10$^{-7}$ and approaching the 10$^{-8}$ s$^{-1}$cm$^{-2}$keV$^{-1}$ level, as required for the future helioscope (Irastorza {\em et al.\/} \cite{ngah}).

Complementary with this experimentation, a detailed simulation of the detector setup has been done. Apart from a Geant4-based geometry description for particle transport of the detector and its shielding, a complete modelization of the detector physics has been implemented. It includes most aspects of the response of the detector and the electronics chain as is currently used in CAST, like electron diffusion along the drift path, signal generation and pixelization in the readout, and temporal pulse shaping by the front-end electronics. The simulated signals reproduce remarkably well the experimental data, in detector-specific aspects like multiplicity (number of strips typically triggered by an event) or the temporal information of the mesh signal (Tom\'as {\em et al.\/} \cite{alfredo}). The offline cuts applied to the real data are equally well applied to the simulated data, reproducing the background rejection factors obtained experimentally. This simulation is helping us to understand the relevant background population after the rejection mechanisms and it is being used to create a detailed background model of the detector in order to eventually identify and reduce effective sources of background.

\section{Micromegas for double beta decay}

The neutrinoless double beta decay of $^{136}$Xe could be searched for by a calorimetric gaseous TPC. This approach was pioneered by the Gothard TPC in the 90's, although only recently a competitive implementation of a gaseous TPC is considered feasible thanks to the latest advances in TPC readouts. The NEXT (Neutrino Xenon TPC) experiment is considering a 100 kg gas Xe TPC for this goal, to be operated at the LSC (Granena {\em et al.\/} \cite{loi}). Here we describe the latest results performed in the context of the NEXT project to develop Micromegas readouts for $\beta\beta$ searches (Cebri\'an {\em et al.\/} \cite{Ceb10}). Although the decided baseline for the NEXT100 detector is an electroluminescent photosensor readout, the development of Micromegas is still motivated as a backup option or for eventual future extension to larger masses, due to the promising prospects for large areas offered by MPGDs. Another collaboration, EXO-gas, also develops a gas TPC to search for the double beta decay of $^{136}$Xe, and considers Micromegas as an option for its readout (Franco {\em et al.\/} \cite{exomm}).

The strong points of using a gas TPC with respect to competing technologies reside in the potentially very good
energy resolution which can be achieved, while taking advantage of the background rejection power provided by the topological
information of the electron tracks. The requirements on the readout include ability to operate at high pressure, sufficient granularity (topological information), good energy resolution and low radioactivity. Micromegas of the microbulk type were identified as promising options because of the best energy resolutions obtained among MPGDs, the well proven capability for a granular design and the fact that they are manufactured out of kapton and copper foils, two materials of known radiopurity. Indeed, the radiopurity of several samples (from raw, to fully manufactured ones) was studied with a high purity Ge detector at the LSC. The results obtained show radioactivity levels below 30$\,\mu$Bq/cm$^2$ for Th and U chains
and $\sim$60$\mu$Bq/cm$^2$ for $^{40}$K (Cebri\'an {\em et al.\/} \cite{Ceb10b}). These results are comparable to the cleanest detector components of the most stringent low background experiments at present, despite the fact that the readouts measured were manufactured
without any special care from the radiopurity point of view. In any case, work is ongoing to increase the sensitivity of the measurements, and further reduce any possible remaining radioactivity.

\begin{figure}[htb!]
\begin{center}
\includegraphics[height=12pc]{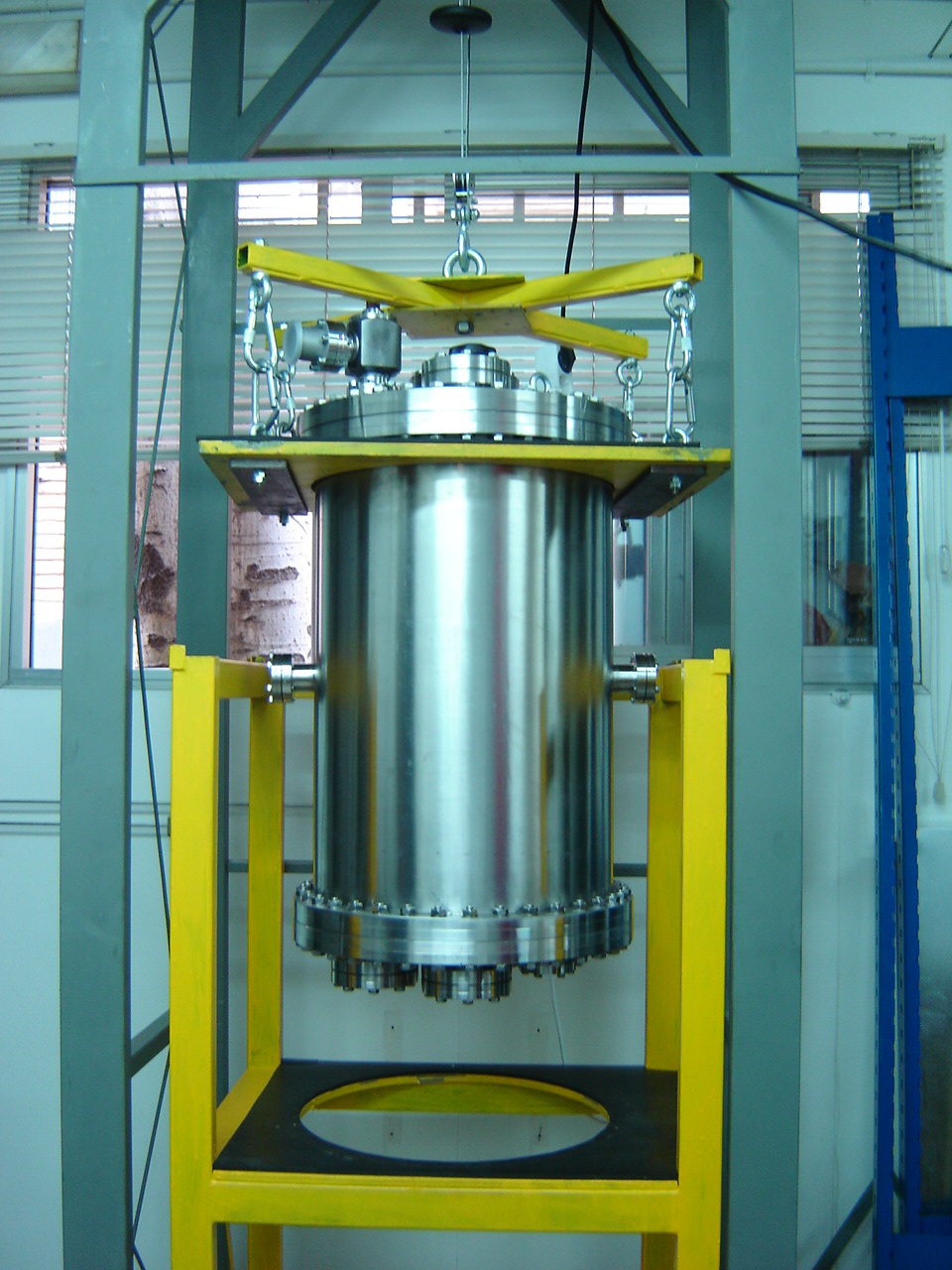}
\includegraphics[height=10pc]{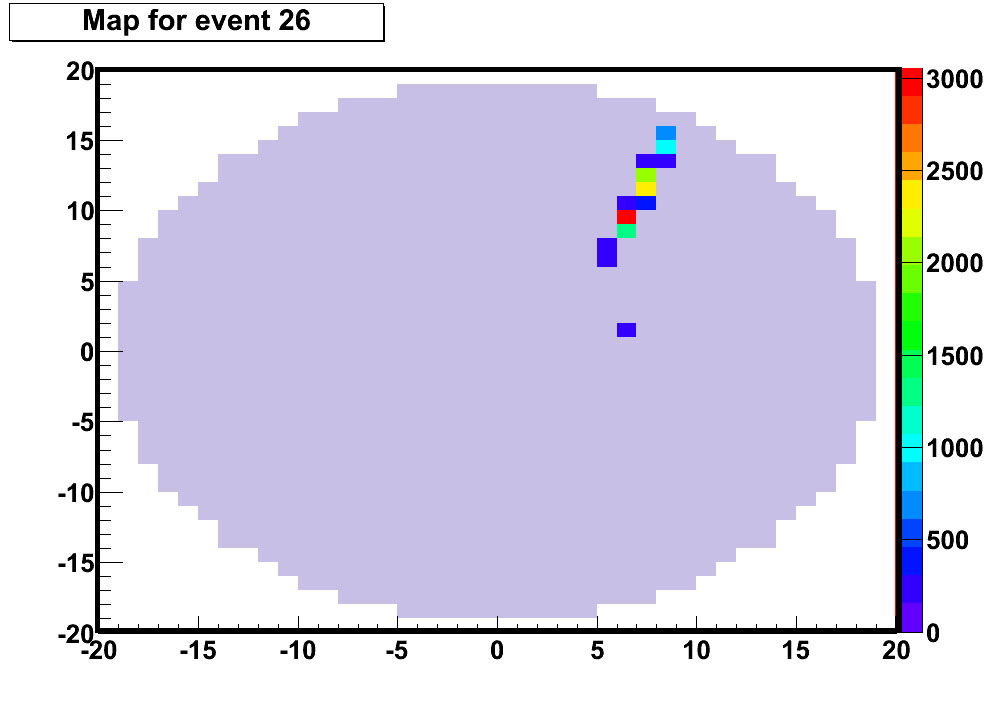}
\caption{\label{next1} Left: a photo of the NEXT-1-MM prototype and its supporting structure. Right: an alpha track of approximately 4$\,$cm obtained with the $^{222}$Rn source in NEXT-1-MM. The shaded area corresponds to the active surface of the bulk detector}
\end{center}
\end{figure}

The work performed up to now has been focused in establishing the capability of microbulk readouts to work in high pressure Xe, and more specifically to measure their energy resolution in those conditions. For that task two prototypes have been built. The first one, known as NEXT-0-MM, is a stainless steel vessel of 2$\,$litres, with a diameter of approx. 14$\,$cm and a drift region of 6$\,$cm, and it is devoted to measurements with small scale readouts, to study gain, operation point, and energy resolution with low energy gammas or alphas. The second prototype, of much larger size (drift of 35 cm and a readout area of 30 cm diameter), NEXT-1-MM, is capable of fully confining a high energy electron track and will therefore probe the detection principle in realistic conditions. A picture of this detector is shown in figure \ref{next1}. We want to stress that work with high pressure and high purity Xe implies very stringent restrictions on leak-tightness and outgassing in the chambers, as well as special gas distribution system with recirculation and filtering stages. Both prototypes have been built keeping combined specifications on ultra high vaccum and high pressure, and keeping a strict control of the materials to be installed inside (outgassing measurements, bake-out). A description of both prototypes can be found elsewhere (G\'omez \etal\ \cite{hector}, Dafni {\em et al.\/} \cite{Dafni:2011zz}), as well as the various tests performed in NEXT-0-MM with small ($\sim 3$ cm diameter) readouts, and with a first pixelised 10$\times$10$\,$cm$^2$ (the largest at the time) microbulk readout that allowed to register the first alpha tracks. Only a summary of the latest results will be given here.

\begin{figure}[htb!]
\begin{center}
\includegraphics[height=9pc]{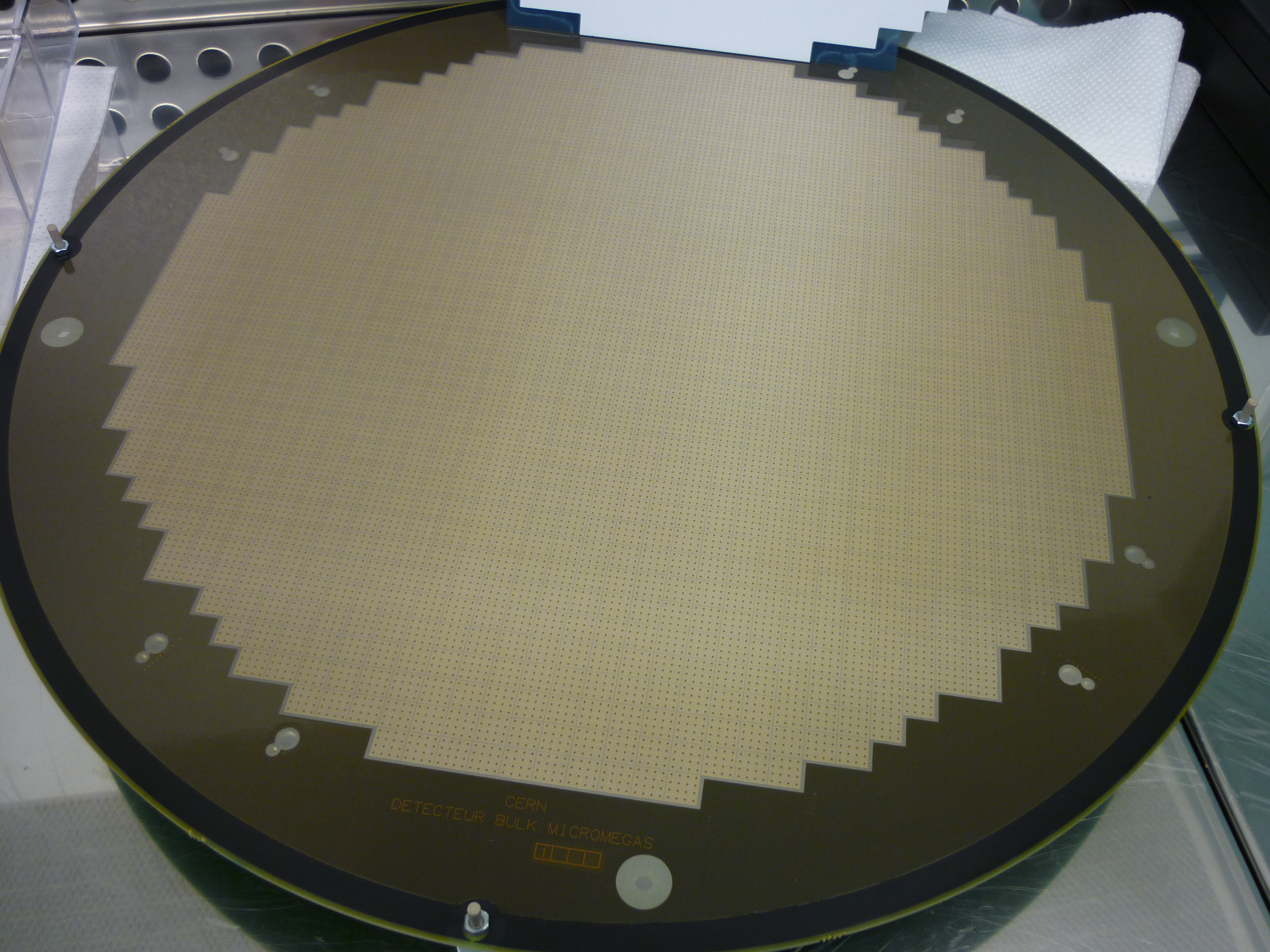}
\includegraphics[height=9pc]{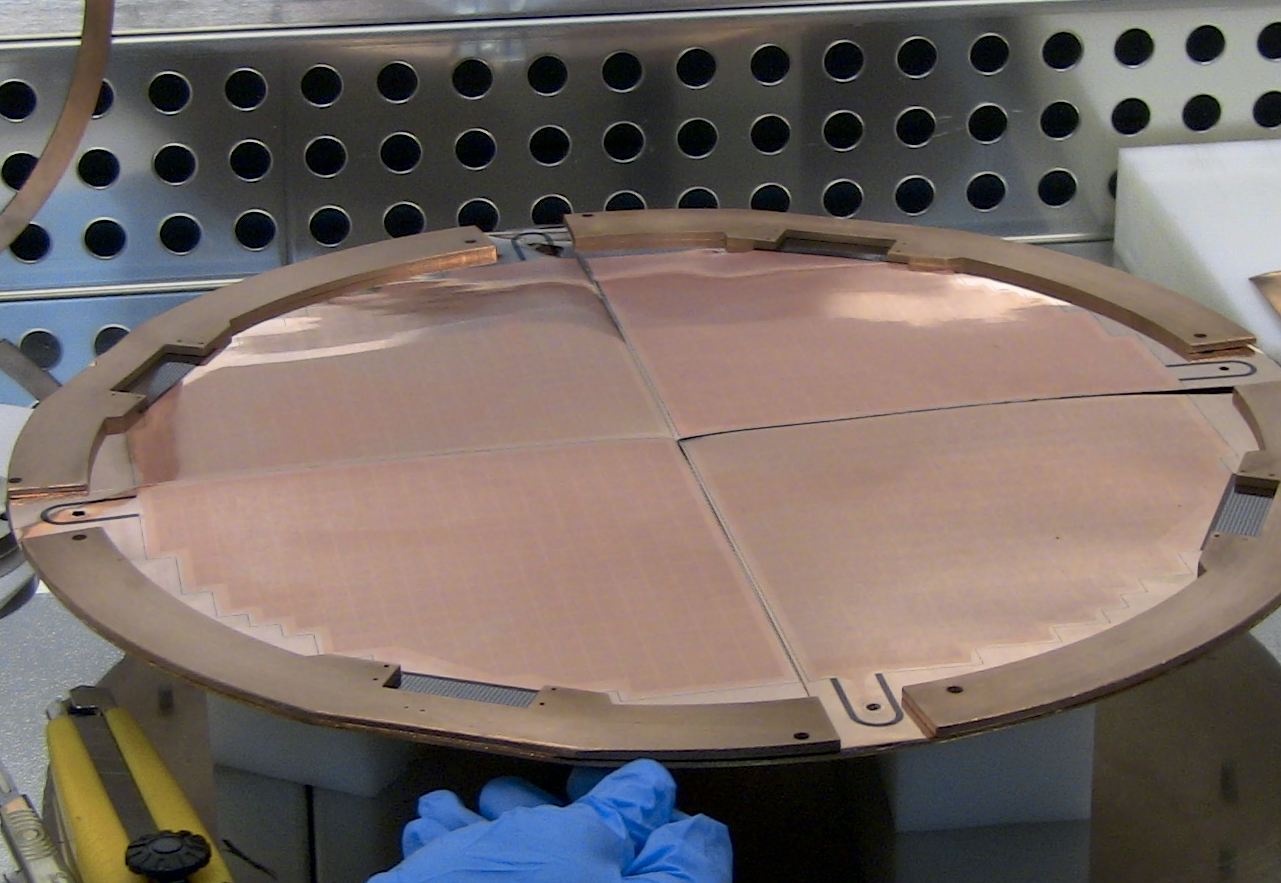}
\caption{\label{next1readout} On the left a photo of the bulk Micromegas detector first installed in the NEXT-1-MM. It has a diameter
of 30$\,$cm and its anode is segmented to $\sim$1200 pixels. On the right the new microbulk version of the same readout, recently installed in the detector. It is divided in 4 circular segments, and constitutes the largest surface built up to now of this type of readout.}
\end{center}
\end{figure}

First tests were performed in Argon-iC$_4$H$_{10}$ mixtures, and later on in pure Argon
and in pure Xenon. Data have been taken in pressures up to $\sim8$ bar. The first remarkable results is that microbulk Micromegas work well in pure high pressure Xe reaching gains above 100. This is remarkable because in absence of quencher early breakdown occurs in all other MPGDs (Balan {\em et al.\/} \cite{Bal11}).
For the 5.5$\,$MeV alpha peak of the $^{241}$Am, energy resolutions down to $\sim$ 2 \% FWHM in pure Ar or Xe, roughly independent of pressure, have been measured (to be compared with 0.7\% FWHM in Ar-2\%iC$_4$H$_{10}$ at 4.75 bar (Dafni {\em et al.\/} \cite{Daf09})). For the low energy 59.5$\,$keV photon peak of the $^{241}$Am (visible when blocking the alpha emission) an energy resolution of 7.8\% FWHM at 2$\,$bar and 9.3\% FWHM at 3.5$\,$bar have been achieved. Data taken in a different setup (Balan {\em et al.\/} \cite{Bal11}) is consistent with this values.

Although the operation in pure Xe is a remarkable achievement for a MPGD readout, the possibility of adding a quencher to the Xe may bring further advantages. We have tested several mixtures, being the most promising Xe+TMA. TMA (trimethilamine) forms a Penning mixture with Xe and indeed much higher gains are observed for the same voltage with respect to pure Xe. Preliminary results point also to energy resolutions almost a factor $\sim2$ better. A publication is being prepared on our tests with this mixture.

The larger prototype NEXT-1-MM was first equipped with a more conventional bulk Micromegas readout, shown in figure \ref{next1readout} with 1252 pixels independently read. Electronics based on the AFTER chip was used for the data acquisition. First alpha tracks were recorded with NEXT-1-MM using this readout (see figure \ref{next1}). Recently the definitive microbulk version of this readout (figure \ref{next1readout}) have been mounted and tested. It is composed of four circular segments covering the 30 cm diameter circular area. Together they are the largest readout manufactured to date using the microbulk technique. The detector with the new readout is being commissioned at the moment.

\section{Conclusion}
Under the R\&D project T-REX, microbulk Micromegas readouts are being developed for their use in rare event searches, under the light of ultra-low background techniques, like shielding, radiopurity of materials or offline cuts using the topology information provided by the readout. Latest advances in connection with two rare event applications, solar axions and double beta decay, have been presented.

\section{Acknowledgements}
We want to thank our colleagues of the groups of the University of Zaragoza, CEA/Saclay and our colleagues from the
CAST, NEXT and RD-51 collaborations for helpful discussions and encouragement.
We thank R. de Oliveira and his team at CERN for the manufacturing of the microbulk
readouts. We acknowledge support from the European Commission under the European Research Council T-REX Starting
Grant ref. ERC-2009-StG-240054 of the IDEAS program of the 7th EU Framework Program. We also acknowledge support
from the Spanish Ministry of Science and Innovation (MICINN) under contract ref. FPA2008-03456, as well as under
the CUP project ref. CSD2008-00037 and the CPAN project ref. CSD2007-00042 from the Consolider-Ingenio 2010
program of the MICINN. Part of these grants are funded by the European Regional Development Fund (ERDF/FEDER).


\end{document}